\documentclass[%
twocolumn,
 amsmath,amssymb,
 aps,
 prc,
floatfix,
]{revtex4-1}
\usepackage{color}
\usepackage{graphicx}
\usepackage{dcolumn}
\usepackage{bm}
\usepackage{natbib}

\begin{document}

\preprint{APS/123-QED}

\title{Role of hexadecapole deformation of projectile $^{28}$Si in heavy-ion fusion reactions\\ near the Coulomb barrier}

\author{Gurpreet Kaur$^{1}$}
\author{K. Hagino$^{2,3}$}
\author{N. Rowley$^{4}$}

\affiliation{$^1$Department of Physics, Panjab University, Chandigarh-160014, INDIA}
\affiliation{$^2$Department of Physics, Tohoku University, Sendai 980-8578, Japan}
\affiliation{$^3$Research Center for Electron Photon Science, Tohoku University, 1-2-1 Mikamine, Sendai 982-0826, Japan}
\affiliation{$^4$Institut de Physique Nucl\'eaire, UMR 8608, CNRS-IN2P3}
\affiliation{Universit\'e de Paris Sud, 91406 Orsay Cedex, France}

\begin{abstract}

The vast knowledge of strong 
influence of quadrupole deformation $\beta_2$ of colliding nuclei 
on heavy-ion subbarrier fusion reactions 
inspires a desire to quest the sensitivity of fusion dynamics 
to higher order deformations, such as $\beta_4$ and $\beta_6$ deformations. 
However, such studies have rarely been carried out, 
especially for deformation of projectile nuclei. 
In this article, we investigated the role of $\beta_4$ of 
the projectile nucleus in fusion of the $^{28}$Si + $^{92}$Zr system. 
We demonstrated that the fusion barrier distribution is sensitive to the 
sign and the value of the $\beta_4$ parameter of the projectile, $^{28}$Si, 
and confirmed that the $^{28}$Si nucleus has a large positive $\beta_4$. 
This study opens an indirect way to estimate deformation parameters 
of radioactive nuclei using fusion reactions, which is otherwise 
difficult due to experimental constraints.
\end{abstract}

          
\date{\today} 

\maketitle

\section{\label{sec:level1}{INTRODUCTION}}

Gaining insight into the role of nuclear intrinsic degrees of 
freedom in heavy-ion fusion reactions has been a motivation of many 
experimental and theoretical studies in the current nuclear 
research \cite{Steadman86,BT98,DHRS98,HT12,Back14,MS17,Andres88}. 
During the fusion process, the nuclear intrinsic degrees of freedom, 
such as inelastic excitations, neutron transfers, static or dynamical 
deformation, are coupled to the relative motion of the interacting nuclei, 
and significantly affect the fusion dynamics. 
Experimental signatures of these couplings have been observed 
via a subbarrier fusion enhancement of fusion cross sections 
and a deviation of fusion barrier 
distributions from a simple one-peaked function 
\cite{Steadman86,BT98,DHRS98,HT12,Back14,MS17}. 
Comparisons of these experimental data with coupled-channels 
calculations have established the role of various 
couplings in heavy-ion fusion mechanism \cite{BT98,HT12}. 

An important question to be addressed is what are the relevant 
degrees of freedom one has to consider in a description of 
the fusion dynamics. For deformed nuclei, 
the role of quadrupole deformation $\beta_2$ of the colliding nuclei 
in fusion is significant and has been well established \cite{DHRS98}. 
With the increasing experimental knowledge on the 
role of quadrupole collectivity 
in fusion, the sensitivity to the hexadecapole deformation $\beta_4$ is 
next to explore. 
In this connection, a measurement by Lemmon {\it et al.} \cite{LemmonPLB93} for $^{16}$O+$^{154}$Sm and 
$^{16}$O+$^{186}$W fusion reactions 
has clearly shown the sensitivity of 
fusion barrier distributions to the sign of $\beta_4$ of the target 
nuclei (see also Refs. \cite{Leigh95,TLD95}). 
The effect of the $\beta_6$ (hexacontatetrapole) deformation, has also been investigated in Refs. \cite{RHT00,Morton01}. 

A study of $\beta_4$ is significantly important 
also in connection to its association with the synthesis of superheavy 
elements (SHEs). 
That is, 
the hexadecapole deformation may significantly affect the height of fusion 
barrier, which in turn influences the fusion probability, thus the 
formation probability of SHEs \cite{IM96}. 
It has theoretically been argued that a $\beta_4$ deformation may 
help fusion (both hot and cold fusion reactions) leading to SHEs, 
depending on the choice of the reaction partners \cite{Gupta05}. 

In this respect, an interesting observation has appeared recently 
while investigating the experimental fusion barrier distribution 
for the $^{28}$Si + $^{154}$Sm system \cite{gkPRC-2016}. 
In this experiment, 
the barrier distribution was extracted using 
quasi-elastic back-scattering \cite{TLD95,HR04}. 
Despite the well-established rotational nature of $^{28}$Si 
(having both quadrupole and hexadecapole deformations), 
it was found that a coupled-channels calculation with a vibrational 
coupling to its first 2$^+$ state reproduces the structure of 
the barrier distribution 
rather well. Subsequently, it was observed that the resolution of 
this anomaly lies in the large hexadecapole deformation parameter 
of $^{28}$Si, which has the opposite sign to the quadrupole 
deformation parameter. 
That is, the contribution to the reorientation coupling ($2_1^+\rightarrow2_1^+$) 
from the quadrupole deformation is largely canceled out by 
that from the hexadecapole deformation, making the rotational coupling scheme look like the vibrational 
coupling scheme for this system. 
This leads to almost identical results for the two coupling schemes. 
Since the quasi-elastic backward scattering is a process complementary 
to fusion, it thus shows a sensitivity of fusion mechanism to 
the hexadecapole deformation of $^{28}$Si.

In Ref. \cite{Newton2001}, 
Newton {\it et al.} 
studied the experimental fusion barrier distribution for 
the $^{28}$Si + $^{92}$Zr system, and 
reached the same conclusion as in Ref. \cite{gkPRC-2016} 
for the $^{28}$Si + $^{154}$Sm system. 
That is, the authors of Ref. \cite{Newton2001}  
have reported that 
treating the 2$^+$ state in $^{28}$Si as a phonon state rather 
than a rotational state with oblate deformation 
gives a somewhat better fit to the experimental fusion 
barrier distribution. 
Moreover, treating the $^{28}$Si nucleus as a prolate rotor leads to 
a poor representation of the data. 
They have argued 
that there is not strong evidence from the fusion data to 
distinguish between $^{28}$Si being a vibrational nucleus 
or an oblate deformed nucleus. 

The aim of this paper is to reanalyse the fusion 
barrier distribution for the 
$^{28}$Si + $^{92}$Zr system which Newton {\it et al.} have studied, 
and to clarify the role of hexadecapole 
deformation of the $^{28}$Si nucleus. 
We shall show that a large positive value for $\beta_4$ 
leads to fusion barrier distributions calculated with the rotational 
coupling scheme which look similar to those with the vibrational scheme. This result cannot be regarded as a direct measurement of $\beta_4$, but it strongly suggests that $^{28}$Si is a deformed nucleus with a large positive hexadecapole parameter, $\beta_4$.

\section{Coupled-channels calculations for $^{28}$S\lowercase{i} 
+ $^{92}$Z\lowercase{r} system}

\begin{figure}
\centering
\includegraphics[width=8.0cm,height=7.0cm]{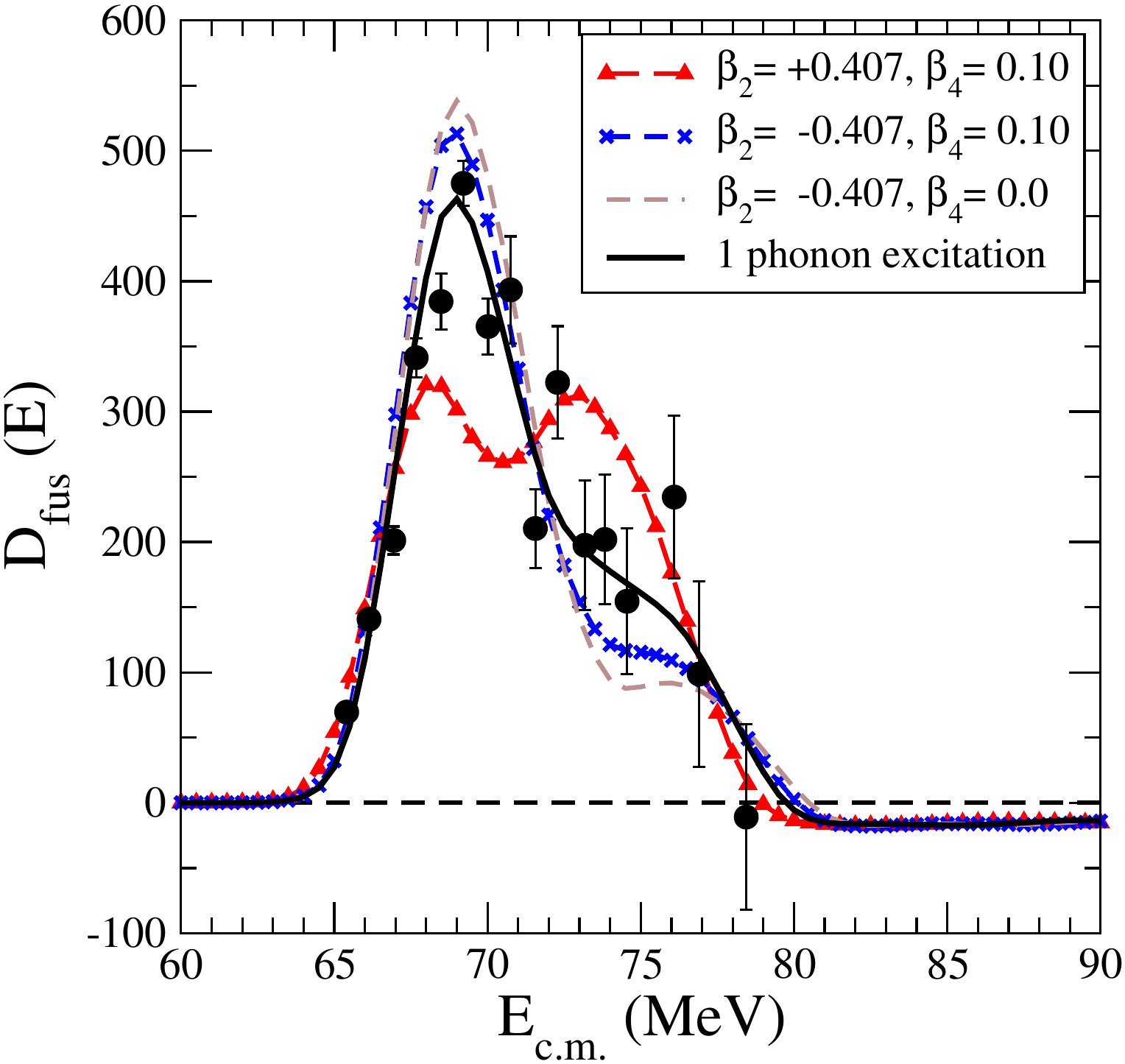}
\caption{\label{fig-1} 
A comparison of fusion barrier distributions for the 
$^{28}$Si + $^{92}$Zr system obtained with several coupling schemes 
for the coupled-channels calculations. 
The solid line shows the result with the vibrational coupling to 
the first 2$^+$ state of $^{28}$Si, 
along with the vibrational excitation of $^{92}$Zr. 
The dashed line shows the result of the rotational couplings to the 
2$^+$ state of $^{28}$Si with deformation parameters of 
$\beta_2=-0.407$ and $\beta_4=0.0$. 
On the other hand, the dashed lines with 
triangles and crosses are obtained with 
$\beta_2=+0.407$ and $\beta_2=-0.407$, respectively, together with 
$\beta_4=0.1$. 
The experimental data, taken from Ref. \cite{Newton2001}, 
are shown with filled circles.}
\end{figure}

To clarify the influence of hexadecapole deformation of $^{28}$Si on 
the fusion of $^{28}$Si + $^{92}$Zr system, we have performed the 
coupled-channels calculations using the computer code {\tt CCFULL} 
\cite{HRK99}. 
To this end, we have used a Woods-Saxon potential, whose diffuseness parameter was fixed to be $a_0$=1.03 fm. Notice that a large value of diffuseness parameter has been found to reproduce high-precision fusion cross sections in many systems \cite{Newton2001}. The exact origin of this phenomenon has not been clarified, and the phenomenon has been referred to as the surface diffuseness anomaly. Here, we follow Ref. \cite{Newton2001}
and take $a_0$=1.03 fm. We have checked that the agreement of the calculation with the experimental data becomes worse if we use a smaller value of $a_0$, such as $a_0$=0.7 fm.
Notice that results are almost independent of the precise values of $V_0$ and $R_0$ as long as the barrier height is reproduced. For excitations in the target nucleus, $^{92}$Zr, we have included a coupling to the one quadrupole phonon state at 0.934 MeV with the deformation parameter of 0.13. 

The dashed line in Fig. \ref{fig-1} shows 
the fusion barrier distribution for the $^{28}$Si + $^{92}$Zr system 
when the coupling to the 2$^+$ state in $^{28}$Si is included assuming 
an oblate rotor with $\beta_2=-0.407$ {\cite{Raman} and $\beta_4$=0. 
Here, the fusion barrier distribution is defined as \cite{RSS91}, 
\begin{equation}
D_{\rm fus}(E)=\frac{d^2(E\sigma_{\rm fus})}{dE^2},
\end{equation}
where $E$ is the incident energy in the center of mass frame and 
$\sigma_{\rm fus}$ is a fusion cross section. 
The experimental fusion barrier distribution 
was extracted with a point difference formula 
with $\Delta E\sim$ 2 MeV \cite{Newton2001}, 
and the same procedure was applied to the theoretical fusion barrier 
distribution as well. 
In the figure, one can find that this calculation captures the main 
structure of the barrier distribution, but the experimental data 
around $E_{\rm c.m}=75$ MeV are not well accounted for. 
The calculation is somewhat improved by taking into account a finite 
value of $\beta_4$, e.g., $\beta_4$=+0.10, the value which was employed 
in Ref. \cite{Danu14}, as is shown by the dashed line with crosses. 
On the other hand, when the quadrupole deformation of $^{28}$Si was 
taken to be positive, 
the shape of fusion barrier distribution becomes inconsistent 
with the experimental data (see the dashed line with 
triangles), 
supporting an oblate deformation of $^{28}$Si \cite{SF80}. 
The solid line in the figure shows the result with 
the vibrational excitation $^{28}$Si, in which the first 2$^+$ state 
is treated as a one phonon state in the harmonic oscillator approximation. 
One can clearly see that this calculation better reproduces the experimental 
fusion barrier distribution, compared to the rotational 
coupling with $\beta_4$=0.10, as has been 
pointed out in Ref. \cite{Newton2001}.

\begin{figure}
\centering
\includegraphics[width=8.0cm,height=7.0cm]{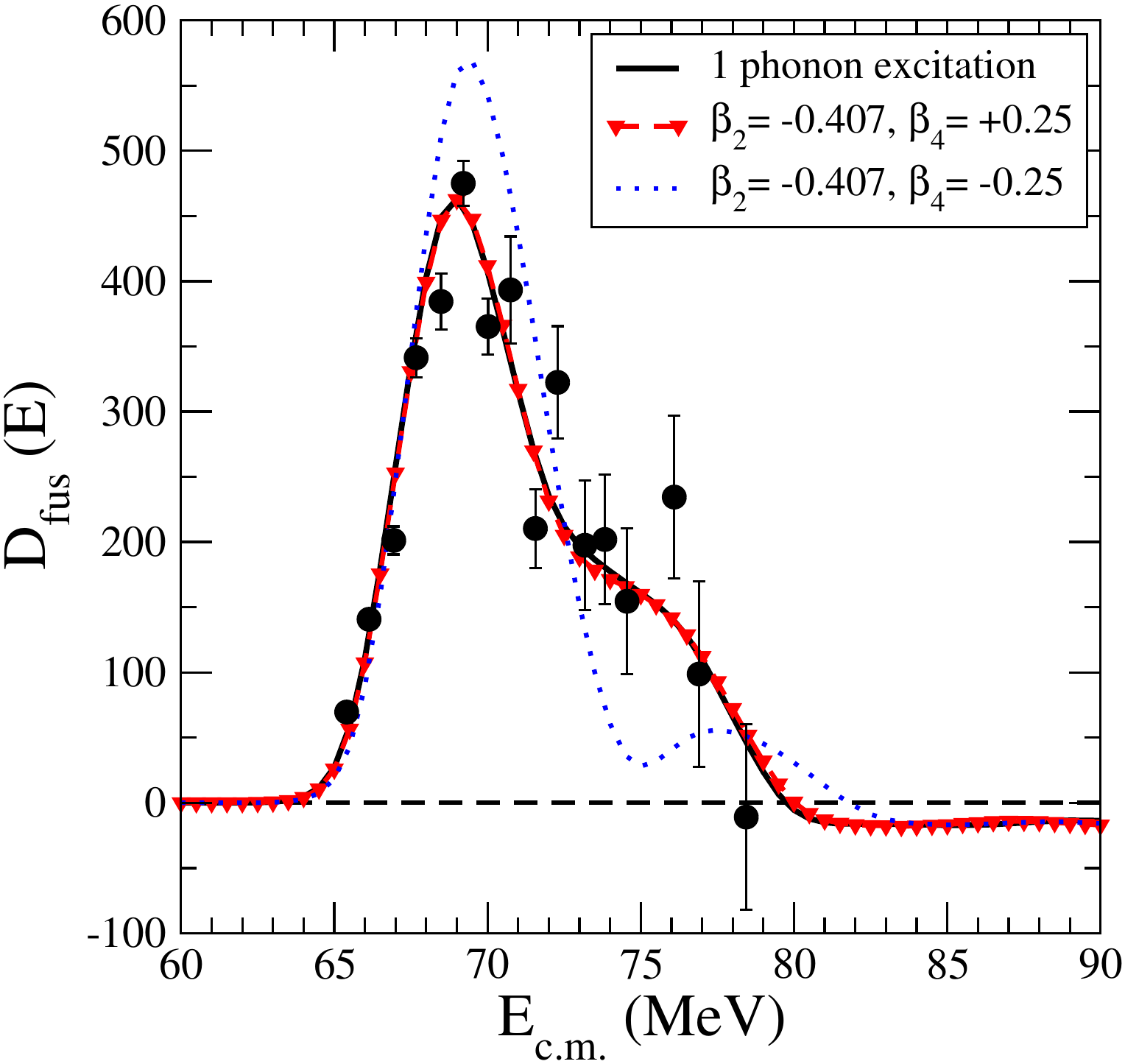}
\caption{\label{fig-2} 
The fusion barrier distribution for the $^{28}$Si + $^{92}$Zr system 
obtained with the rotational coupling scheme for $^{28}$Si 
with $\beta_2=-0.407$ and $\beta_4=0.25$ 
(the dashed line with triangles) and with $\beta_2=-0.407$ and $\beta_4=-0.25$ (the dotted line). The meaning of the solid line is the same 
as in Fig. \ref{fig-1}. 
}
\end{figure}

In order to see the sensitivity of the results to $\beta_4$ in 
the rotational coupling scheme, the 
dashed line with 
triangles in Fig. \ref{fig-2} show the barrier distribution obtained 
with a larger value of $\beta_4$, that is, $\beta_4=0.25$. 
This is the value obtained by 
M\"oller and Nix \cite{Moller95} 
by using 
the finite-range droplet model with 
spherical-harmonic expansions. This value is also consistent with the one 
obtained with proton scattering experiments, i.e., +0.25$\pm$0.08 \cite{Swiniarski69}. Earlier experiments for electron scattering \cite{electron}, neutron scattering \cite{neutron,Haouat1984} and alpha particle scattering \cite{alpha} indicate that the value of $\beta_4$ in $^{28}$Si is +0.10, +0.18$\pm$0.02/+0.20$\pm$0.05 and +0.08$\pm$0.01, respectively. Although these values are somewhat different from each other, all of these values point to a large
value of $\beta_4$.
Interestingly, the rotational calculation with $\beta_4=0.25$ yields an almost identical result to the result of the vibrational coupling scheme shown by the solid line in the figure. This is in the same situation as in the $^{28}$Si+$^{154}$Sm system discussed in Ref. \cite{gkPRC-2016}.

Within the space of the ground state (0$^+$) and the first 2$^+$ state, 
the difference between the rotational and the harmonic vibrational coupling 
schemes is found only in the re-orientation term. 
That is, there is no coupling from the $2^+$ state to the same state, 
$2^+$, in the vibrational coupling, while this coupling is finite in 
the rotational coupling (compare between Eqs. (3.41) and (3.49) 
in Ref. \cite{HT12}). 
It is important to notice here that the 2$^+$ state is coupled to 
itself by both the quadrupole and the hexadecapole deformations. 
In fact, the reorientation term is given by 
(see Eq. (3.58) in Ref. \cite{HT12}),
\begin{eqnarray}
O_{22}&=&\langle Y_{20}|\beta_2R_PY_{20}(\theta)+\beta_4 R_P Y_{40}(\theta)
|Y_{20}\rangle, \\
&=&
\frac{5\sqrt{5}}{\sqrt{4\pi}}\beta_2 R_P
\left(\begin{array}{ccc}
2&2&2\\
0&0&0
\end{array}\right)^2 +
\frac{15}{\sqrt{4\pi}}\beta_4 R_P
\left(\begin{array}{ccc}
2&4&2\\
0&0&0
\end{array}\right)^2, \nonumber \\
\label{eq:reorientation}
\end{eqnarray}
where $R_P$ is the radius of the projectile 
nucleus and the 3$j$ symbols read,
\begin{equation}
\left(\begin{array}{ccc}
2&2&2\\
0&0&0
\end{array}\right)=-\frac{2}{\sqrt{70}},
~~~
\left(\begin{array}{ccc}
2&4&2\\
0&0&0
\end{array}\right)=+\frac{2}{\sqrt{70}}.
\end{equation}
With $\beta_2=-0.407$ and $\beta_4=+0.25$, the first and the second terms 
in Eq. (\ref{eq:reorientation}) read 
$-0.073R_P$ and $+0.060R_P$, respectively, which are largely canceled 
with each other, leading to the situation which is close to the 
vibrational coupling scheme (the perfect cancellation is achieved for 
$\beta_4/\beta_2=-\sqrt{5}/3=-0.745$). As a matter of fact, the similarity disappears when we take $\beta_4=-0.25$, as shown by the dotted line in Fig. \ref{fig-2}.
Therefore, 
even though the 
result with the vibrational coupling scheme 
may lead to a good reproduction of the experimental data, 
this does not imply that 
$^{28}$Si is a vibrational spherical nucleus. 
A nice reproduction is simply 
due to an accidental cancellation of the 
reorientation term originated from 
the large value of $\beta_4$, 
and the rotational excitation of the $^{28}$Si projectile 
still plays an important role in the fusion of this nucleus. 

\begin{figure}
\centering
\includegraphics[width=8.0cm,height=7.0cm]{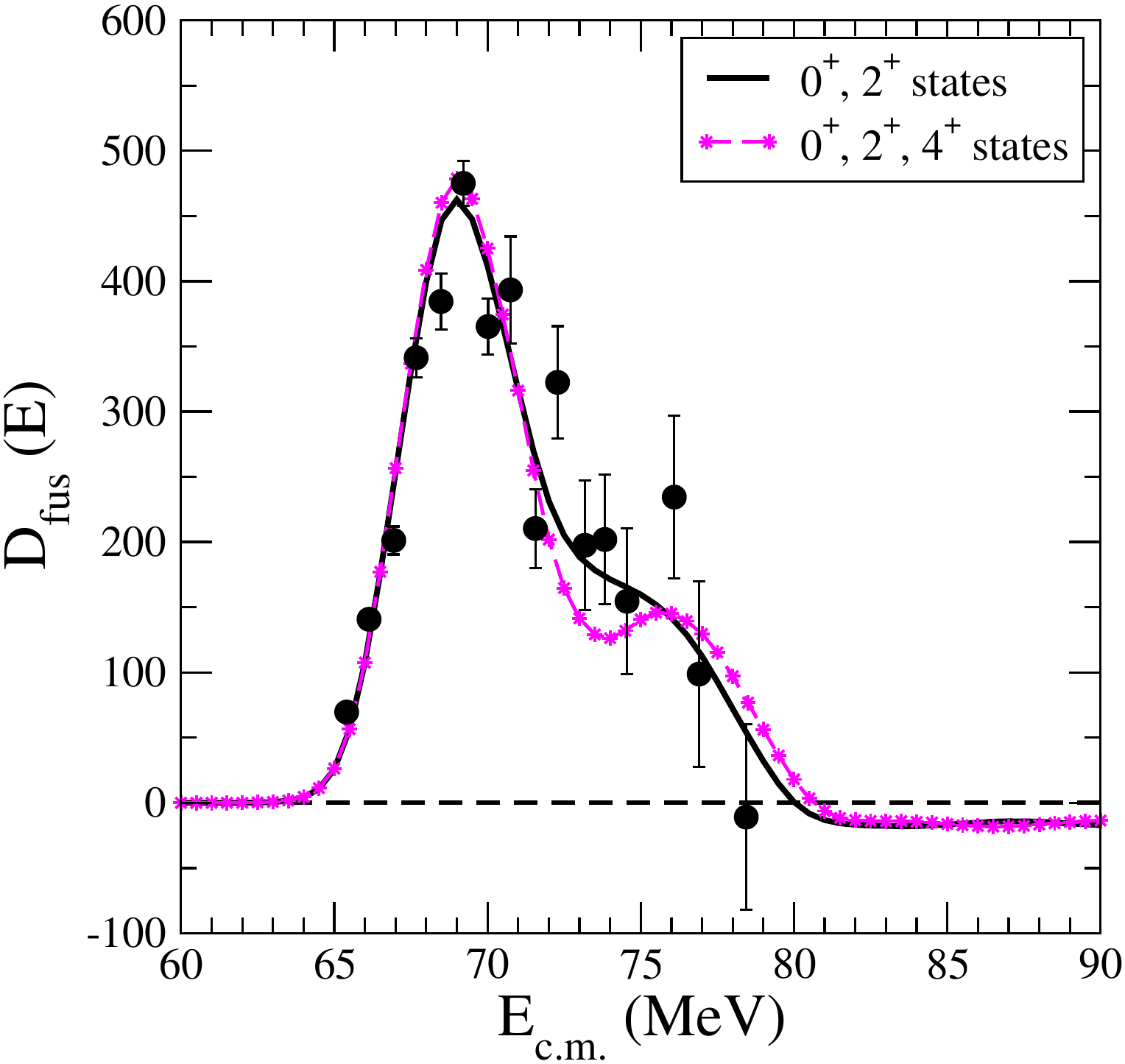}
\caption{\label{fig-3} A comparison of fusion barrier distributions 
for the $^{28}$Si + $^{92}$Zr system obtained with the rotational 
coupling scheme with a truncation of the 
ground state rotational band of $^{28}$Si at the 2$^+$ state (the 
solid line) and at the 4$^+$ state (the dashed line with stars) in the 
coupled-channels calculations. 
}
\end{figure}

A large hexadecapole deformation of $^{28}$Si should accompany 
a strong direct coupling from the ground state to the $4^+$ state. 
The 4$^+$ state also couples to the $2^+$ state with both the quadrupole 
and the hexadecapole terms (notice that there is no hexadecapole coupling 
between the 0$^+$ state and the 2$^+$ state). 
In order to check the influence of the 4$^+$ state, 
the dashed line with stars 
in Fig. \ref{fig-3} shows the result obtained by 
including the ground state rotational band of $^{28}$Si up to 
the 4$^+$ state with the deformation parameters of $\beta_2=-0.407$ 
and $\beta_4=+0.25$. 
The inclusion of the 4$^+$ state somewhat perturbs the shape of 
fusion barrier distribution, and the agreement with the experimental 
data is slightly worsened. However, the calculated fusion barrier 
distribution is still within the error bars of the experimental distribution 
and there remains a similarity to the barrier distribution for 
the vibrational coupling scheme. 
We have confirmed that the agreement is not significantly improved 
even with a larger value of $\beta_4$, that is, $\beta_4=0.30$.
We have also checked the influence of the octupole excitation to the
3$^-$ state at 6.878 MeV in $^{28}$Si and have confirmed that the inclusion
of this state simply shifts the barrier distribution in energy
by $\approx$ 1.5 MeV without significantly changing its shape. As has been pointed out e.g., in
\cite{HT12}, excitation to a state with large excitation energy, such
as the 3$^-$ state in $^{28}$Si, simply lead to a renormalization of the
fusion barrier, thus do not significantly influence the fusion dynamics.
We have also found that 
the results converge rapidly on adding the higher members in 
the rotational band, beyond $4^+$, of $^{28}$Si due to the finite excitation 
energy. This latter fact is another 
necessary condition to have a similarity between 
the rotational coupling and the vibrational coupling schemes. That is, when 
higher members in the ground state rotational band contribute significantly 
to the fusion dynamics, 
which is typically the case for fusion of medium-heavy nuclei such 
as $^{16}$O+$^{154}$Sm, 
the resultant fusion barrier distribution 
differs considerably from fusion barrier distributions for vibrational 
nuclei \cite{DHRS98,HT12,Leigh95}.

\section{Summary and discussions}

In summary, we have carried out the coupled-channels calculations for the 
$^{28}$Si + $^{92}$Zr fusion reaction and 
have shown that the fusion process is sensitive to the hexadecapole 
deformation of the $^{28}$Si nucleus. 
We have demonstrated that the reorientation term for the 2$^+$ state 
is largely canceled out, 
leading to similar results between the rotational and the vibrational 
coupling schemes, even though in reality 
the $^{28}$Si nucleus is not a spherical nucleus. 
This nicely follows the earlier conclusion obtained for 
the $^{28}$Si+$^{154}$Sm reaction \cite{gkPRC-2016}, 
making 
a strong evidence for that 
$^{28}$Si possesses a large positive hexadecapole 
moment.  

In order to have such similarity between results with the rotational 
coupling scheme and those with the vibrational coupling scheme, 
the following two 
conditions are necessary. The first condition is that the quadrupole 
and the hexadecapole deformation parameters have opposite sign to 
each other and the ratio is close to 
$\beta_4/\beta_2=-\sqrt{5}/3=-0.745$. The second condition, which is 
usually satisfied for light deformed nuclei,  
is that the excitation energy of the first 2$^+$ state 
is large so that higher members of the ground state rotational band 
do not significantly contribute. 
The $^{28}$Si nucleus satisfies both 
conditions. In addition to other Si isotopes, another candidate which 
shows the same kind of 
similarity might be $^{38}$Ne. Even though several aspects 
related to the weakly-bound nature of this neutron-rich nucleus 
would have also to be taken into account, this nucleus satisfies the two 
conditions, as the deformation 
parameters for this nucleus 
are predicted to be $\beta_2=-0.302$ and $\beta_4=+0.163$ with 
the FRDM(2012) mass model \cite{FRDM2012} and the energy of the 2$^+$ 
state is predicted to be around 1.05 MeV with a shell model calculation 
\cite{CNP14}. 

The coupled-channels calculations for the $^{28}$Si+$^{92}$Zr
system presented in this paper 
suggest that the fusion mechanism is sensitive to projectile 
excitations. 
This is also relevant to the synthesis of superheavy elements. 
Very recently, barrier distributions were extracted 
using quasi-elastic scattering for reactions to form superheavy 
elements \cite{Tanaka18}. 
Quasi-elastic barrier distributions 
are complimentary to fusion barrier distributions and they have 
smaller error bars on the high energy side. 
It will be an interesting future work to study how the projectile 
excitations influence evaporation residue cross sections for fusion 
reactions of the $^{28}$Si projectile to form superheavy elements. 
Note also that the method based on a 
quasi-elastic barrier distribution will be 
useful to discuss the shape of radioactive nuclei, 
for which the beam intensity is low \cite{HR04}. 

In the past, 
$\alpha$-particle scattering \cite{Lee-PRC75,Baker-NPA76}, electron 
scattering \cite{Cooper-PRC76}, and muonic $x$-rays 
methods \cite{Powers-PRL75} have been used
in order to determine experimentally 
the shape of a deformed nucleus. 
However, $\beta_4$, 
especially its sign,  
is difficult to extract. All the available results for $\beta_4$ 
are model-dependent and quite different from each other 
with large uncertainties. 
As we have discussed in this paper, fusion is sensitive  
not only to the target excitations but also to 
the projectile excitations, and the barrier distribution analysis will offer an alternative powerful method to extract the magnitude and sign of $\beta_4$ for deformed nuclei.

\end{document}